# Phase Tetrastability in Parametric Oscillation

Rubén Martínez-Lorente, Fernando Silva, and Germán J. de Valcárcel*

[1]*Departament d'Òptica, Universitat de València, Dr. Moliner 50, 46100 Burjassot, Spain*
*german.valcarcel@uv.es

The periodic modulation of an oscillator's frequency can lead to so-called parametric oscillations at half the driving frequency, which display bistability between two states whose phases differ by π. Such phase-locking bistability is at the root of the extraordinary importance of parametric oscillation (and amplification) both in fundamental and applied scenarios. Here we put forward a universal method for exciting tetrastability in parametrically-driven systems, which consists in modulating the amplitude of the parametric drive in such a way that its sign alternates periodically in time. This way, multistability can emerge between four states whose phases differ by a multiple of π/2. We prove theoretically the validity of the method, both analytically and numerically, and demonstrate it experimentally in an optical oscillator. The method could be relevant to the fields of pattern formation, (quantum) information processing and simulation, metrology and sensing.

*Introduction.*—The familiar image of a child on a swing extending and bending rhythmically their legs is probably the humbler example of parametric driving, which consists in the periodic modulation of an oscillator's frequency [1-4]. When the frequency of such modulation (call it $2\omega_f$) and the oscillator's bare frequency ($\omega_0$) are in a ratio close to 2:1 (hence $\omega_f \approx \omega_0$), a so-called parametric resonance happens which, above a threshold, leads to oscillations at the subharmonic frequency $\omega_f$. Such oscillations are bistable between two phases that differ by π while, below threshold, phase-sensitive (modulo π) signal amplification is observed around the frequency $\omega_f$. This phase sensitivity is easily understood since the system must be invariant under a shift in time by the modulation period (equal to $\pi/\omega_f$), hence any two oscillatory states of equal amplitude but whose phases differ by π must be equivalent. Parametric oscillation and amplification are observed in very diverse systems, like mechanical/ electromechanical oscillators [5-12], optomechanical cavities [13], granular materials [14], colloidal suspensions [15], fluids [1,4,16-18], Josephson-junction superconducting circuits [19-21], nonlinear optical cavities [22-33], or Bose-Einstein condensates [34-36], to cite some of them. An analogous effect (hence dubbed parametric) occurs as well in degenerate four-wave mixing (FWM) [37,38]: when two strong waves (of frequencies $\omega_\pm$) pump a third-order nonlinear medium, phase-sensitive (modulo π) effects occur at the mean frequency $\omega_0 = (\omega_+ + \omega_-)/2$ [39-41]; this parametric FWM process is produced also when the pumps have equal frequencies [42-49], in which case all four waves are frequency-degenerate and differ in e.g. their propagation directions.

The phase sensitivity of parametric systems is at the root of their relevance in modern applications. On one hand phase-sensitive amplification allows the control over quantum noise, leading to squeezed, entangled, and other quantum states [13,22-24], with their use in sensing, metrology, quantum information, quantum computing and quantum communication [20,25-29]. On the other hand, phase bistability of oscillations allows binary information coding and processing [9,10], including spatially extended [47-49] systems; it also mimics the physics of spin-½ systems, which allows implementing simulators for complex optimization problems, like quantum annealers and coherent Ising machines [11,30-33]; see [50] for a review. In passing it's worth mentioning the role that the parametric instability (also called Faraday instability) has on the generation of supercontinuum and of mode-locked radiation [51-54], and on $\mathcal{PT}$-symmetry physics [55].

In this Letter we show that phase tetrastability can be excited in parametrically-driven systems just by (additionally) modulating periodically the drive amplitude, i.e. using a drive of the form $\lambda(t)\cos(2\omega_f t)$, with $\lambda(t + T) = \lambda(t)$. The only requirements for the amplitude $\lambda(t)$ being: (i) its modulation period must verify $\gamma^{-1} \gg T \gg 2\pi/\omega_f$ being $\gamma$ the system's linear damping, and (ii) its one-period average, $\langle\lambda(t)\rangle$, must be null hence its *sign* must alternate periodically. There are many ways to realize experimentally condition (ii), a simple one being $\lambda(t) = \lambda_0 \cos(2\pi t/T)$, up to an arbitrary (and dynamically unimportant) phase. Note that this modulation in fact corresponds to a two-tone driving with two close frequencies, namely $2\omega_f \pm 2\pi/T$, both close to the second harmonic of the oscillator's bare frequency $\omega_0$.

The present proposal gets inspiration from previous studies on the effect that such a modulation (dubbed rocking [56]) has on the usual resonant, direct driving of oscillators at a frequency close to $\omega_0$ (1:1 direct resonance) [56-59]. It has been shown that the rocking modulation, when relatively fast as compared to the characteristic damping time of the system, transforms the originally phase-monostable system into a phase-bistable one [56-59]. An extended discussion of the physics behind this transformation can be found in [56,59].

We study the effects of the proposed driving method in two nonlinear models, and finally we give experimental evidence of phase tetrastability in an optical oscillator.

*Duffing-type oscillator model*—First, we consider a well-established model for a parametrically-driven mechanical resonator [12,60], namely

$$\ddot{x} + \omega_0^2[1 + \lambda(t)\cos(2\omega_f t)]x + 2\gamma\dot{x} + \alpha x^3 + \eta x^2 \dot{x} = 0, \qquad (1)$$

where $x$ is the oscillator displacement, dots denote differentiation with respect to time $t$, $\gamma$ is the linear damping coefficient, denoted by $\Gamma/2$ in [12] ($\gamma = \omega_0/2Q$, with $Q$ the resonator quality factor), and finally $\alpha$ and $\eta$ are the coefficients of nonlinear dispersion (Duffing type) and dissipation, respectively. In [12], $\lambda$ is a constant as usual in classic parametric driving, while here we assume the form $\lambda(t) = \lambda_0 \cos(2\pi t/T)$. We consider the values [12] $\omega_0/2\pi = 325$ Hz, $Q = 1.8 \times 10^3$, $\alpha = 2.45 \times 10^{10}$ m$^{-2}$ s$^{-2}$, and $\eta = 6.8 \times 10^6$ m$^{-2}$ s$^{-1}$.

In first place we prove analytically that Eq. (1) displays phase-tetrastable oscillations for appropriate values of the parametric drive parameters. We use a standard multiple-scale analysis [61-64], which relies on the existence of disparate time scales in a dynamical system. In our case $\omega_f \approx \omega_0 \gg \gamma$, while the characteristic time scale of the nonlinearities is comparable to the linear damping. In usual parametric driving ($\lambda$ constant), $\lambda = O(\gamma/\omega_0)$, implying that it acts on the same time scale as the rest of terms (but the restoring force) [3,63,64]. Here such scaling for $\lambda$ must be modified because the modulation amplitude is time dependent, on the scale of the modulation period $T$. We find that the sought-for phase tetrastability requires $\lambda = O(\sqrt{\gamma/\omega_0})$ and $\omega_0 \gg T^{-1} \gg \gamma$, which is typical of rocking [56,59]. Finally, we define a normalized detuning as $\theta = (\omega_f - \omega_0)/\gamma$ and assume $\theta = O(1)$, i.e. nearly-resonant driving. Following the standard methods [61-64], an equation for the displacement $x$ is obtained:

$$x(t) = A(t)e^{-i\omega_f t} + c.c. + O(\varepsilon^2), \quad (2a)$$
$$A(t) = \left[e^{L(t)} \operatorname{Re} \psi(t) + ie^{-L(t)} \operatorname{Im} \psi(t)\right]e^{-i\pi/4}, \quad (2b)$$

where $A = O(\varepsilon)$, $\varepsilon \equiv \sqrt{\gamma/\omega_0}$, $L(t) = \xi \sin(2\pi t/T)$, $\xi \equiv \lambda_0 \omega_0 T/8\pi$, and the complex amplitude $\psi$ verifies

$$\frac{1}{\gamma}\frac{d\psi}{dt} = -(1 - i\chi_2 \theta_{\text{eff}})\psi - \frac{1}{2}\chi_2 \tilde{\eta}|\psi|^2 \psi$$
$$- \frac{3}{2}i\kappa\tilde{\alpha}(|\psi|^2\psi + \mu\psi^{*3}), \quad (3a)$$

being $\tilde{\alpha} = \alpha/\omega_0^2$, $\tilde{\eta} = \eta/\omega_0$, $\theta_{\text{eff}} = \theta + \frac{\pi}{16} \frac{\lambda_0}{\gamma T} \frac{I_1(2\xi)}{I_0(2\xi)}$,

$$\chi_k \equiv \langle e^{kL(t)} \rangle, \quad \kappa \equiv \frac{3\chi_4 + 1}{4}, \quad \mu \equiv \frac{\chi_4 - 1}{3\chi_4 + 1}. \quad (3b)$$

In this case $\chi_k = I_0(k\xi)$, with $I_n$ a modified Bessel function of the first kind. Note that $A(t)$ is a slowly-varying complex amplitude, as $L(t)$ and $\psi(t)$ so are: they vary respectively on the scales $T \sim \varepsilon\omega_0^{-1}$, and $\gamma^{-1} \sim \varepsilon^2\omega_0^{-1}$.

The term proportional to $\psi^{*3}$ in Eq. (3a) is responsible for the sought-for phase tetrastability, as it imposes the discrete phase symmetry $S_4: \psi(t) \mapsto \psi(t)e^{i\pi/2}$. This effect requires $\alpha \neq 0$, i.e. a nonlinear dispersion (Duffing type oscillator), while the nonlinear dissipation, governed by $\eta$, is unimportant for that. The steady states of Eq. (3) correspond to phase-locked oscillating states at the frequency $\omega_f$ according to Eq. (2a) and they should be tetrastable according to our analysis.

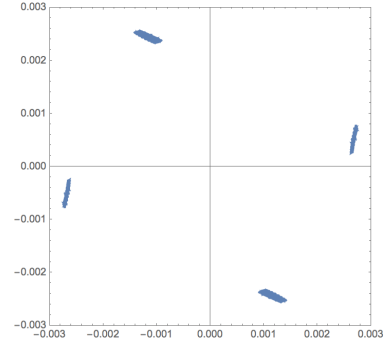

FIG. 1. Phase tetrastability. Four numerical solutions of Eq. (1), with $\lambda(t) = \lambda_0 \cos(2\pi t/T)$, which differ only in their initial conditions. We show a complex-plane representation of the amplitude $A_{\exp}$ given by Eq. (4a). Parameters are [12]: $\omega_0/2\pi = 325$ Hz, $\gamma = 1.14$ s$^{-1}$, $\alpha = 2.45 \times 10^{10}$ m$^{-2}$ s$^{-2}$, and $\eta = 6.8 \times 10^6$ m$^{-2}$ s$^{-1}$. As for the driving, $\omega_f/2\pi = 326.14$ Hz, $\lambda_0 = 0.0335$, and $T = 31$ ms.

We have confirmed this prediction by numerically integrating the original Eq. (1) under a variety of conditions, finding that phase tetrastability is a generic, non-critical phenomenon. As an example, Fig. 1 gives evidence of four stable solutions of Eq. (1) corresponding to the same parameter set, which just differ in their phase by multiples of $\pi/2$. The plot represents an approximation to the amplitude $A$, which we call $A_{\exp}$ and define as

$$A_{\exp}(t) \equiv e^{i(\omega_f t + \varphi)}[x(t) + i\dot{x}(t)/\omega_f], \quad (4a)$$

being $\varphi$ an arbitrary phase (in an experiment, the exponential must be synchronized with the drive). Substituting Eq. (2a) into (4a) and taking into account the slowly-varying character of $A(t)$, we get $A_{\exp}(t) \approx e^{i\varphi}A(t)$. We can further perform an average over one period of the modulation $T$, obtaining

$$\psi_{\exp}(t) \equiv \langle A_{\exp}(t) \rangle \approx \chi_1 e^{i(\varphi - \pi/4)}\psi(t), \quad (4b)$$

where we used Eq. (2b) and $\langle \psi(t) \rangle \approx \psi(t)$. Note that the effect of $\varphi$ is just a rotation on the complex plane.

*Nonlinear Schrödinger-type model*—Next, we consider a universal model for parametrically-excited weakly-nonlinear dispersive waves. Expressing the oscillatory state as $A(\vec{r}, t)e^{-i\omega_f t} + $ c.c, and the drive as $\lambda(t)e^{-i2\omega_f t} + $ c.c., the model can be written as

$$\gamma^{-1}\partial_t A = -A + \lambda(t)A^* + is(|A|^2 - \theta)A + i\nabla^2 A, \quad (5)$$

which is a nonlinear Schrödinger equation, generalized to include, apart from damping, the effect of parametric gain in an explicit way: the term $\lambda(t)A^*$ breaks the continuous phase symmetry of free oscillations down to the discrete one $S_2: A \mapsto e^{i\pi}A$, leading to phase bistability.

Parameter $s = \pm 1$, and accounts for the type of dispersive nonlinearity: self-focusing/weakening ($s = +1$) or defocusing/hardening ($s = -1$) as in the oscillator model (1), while $s$ is introduced in front of the detuning $\theta$ parameter for convenience, $\theta \equiv -s(\omega_f - \omega_0)/\gamma$. Finally, the Laplacian operator $\nabla^2$ has the suitable dimensionality fitting the problem geometry (usually 1D or 2D). All quantities in Eq. (5) are dimensionless but time: the amplitude $A$ has been normalized as to make equal to $s$ the nonlinear dispersion coefficient [equivalent to $\alpha$ in Eq. (1)] which otherwise would appear multiplying the nonlinear term, and space has been normalized to the characteristic dispersion/diffraction length.

Equation (5) is used for studying both extended and localized waves (solitons, breathers and the like) in a huge variety of parametrically pumped systems, including micro- and nano-electromechanical systems [66], fluids [17,18,67], chains of coupled pendula [6,68] or Josephson junctions [69] (via sine-Gordon equation [70-72]), ferromagnets [69], and optical systems [73-76]. We have not included nonlinear dissipation, neither wavenumber-dependent loss in Eq. (5) as that's a pretty usual scenario, see [77] and because we want to keep the presentation simple enough. In any case we have checked that those processes are not essential for the phase tetrastability phenomenon, like in the previous case of the mechanical oscillator model (1).

In order to give more generality to our study, now we consider any periodic parametric-drive amplitude, i.e. not necessarily harmonic, whose one-period average, $\langle\lambda(t)\rangle$, is null. We keep considering the limit where parametric drive is stronger than, and acts on a faster time scale than, dissipation and nonlinearity, but slower than the base frequency $\omega_f$. This allows a standard multiple-scale analysis similar to the previous one, but instead we give here an alternative, simpler treatment leading to the same results. We write $\lambda(t) = \varepsilon^{-1}m(t/T)$, where $\varepsilon \equiv \gamma T \ll 1$, and $m$ is of order unity. In that limit, the leading-order approximation to Eq. (1) reads $\partial_t A = \gamma\lambda(t)A^*$, which has solutions $A(\vec{r},t) = U(\vec{r})e^{L(t)} + iV(\vec{r})e^{-L(t)}$, where $U$ and $V$ are constant in time, $L(t) \equiv \gamma\int\lambda(t)dt$, and we choose $\langle L\rangle = 0$. The expected main effect of the other, smaller terms in Eq. (5) is to introduce "slow" time variations on $U$ and $V$ [78], hence we write $A(\vec{r},t) = U(\vec{r},t)e^{L(t)} + iV(\vec{r},t)e^{-L(t)} + \mathcal{O}(\varepsilon)$, substitute into Eq. (5), and assume that $U$ and $V$ do not vary appreciably along one period. This way a closed equation for $\psi(\vec{r},t) \equiv [U(\vec{r},t) + iV(\vec{r},t)]$ can be derived at $\mathcal{O}(\varepsilon^0)$, which can be written as

$$\frac{1}{\gamma}\frac{\partial}{\partial t}\psi = -(1 + is\chi_2\theta)\psi + i\chi_2\nabla^2\psi + is\kappa(|\psi|^2\psi + \mu\psi^{*3}). \tag{6}$$

Here $\chi_k \equiv \langle e^{kL(t)}\rangle$, $\kappa \equiv \frac{1}{4}(3\chi_4 + 1)$, and $\mu \equiv \frac{\eta_4 - 1}{3\eta_4 + 1}$, as in Eq. (3b). In order to arrive to Eq. (6) we assumed $\chi_{-k} = \chi_k$, which is a common case and in particular that of sine- and square waveforms, for which $\chi_k = I_0(k\lambda_0\gamma T/2\pi)$ and $\chi_k = \frac{\sinh(k\lambda_0\gamma T/4)}{k\lambda_0\gamma T/4}$, respectively.

Equation (6) coincides with Eq. (3a) in the case where the original models match each other. The spatially-uniform steady states of Eq. (6) represent the phase-locked uniform oscillations of the system and can be written as $\psi = \sqrt{\rho/\kappa}\,e^{i\phi}$, with $\rho$ and $\phi$ real constants. The "intensity" $\rho = \left(\eta_2\theta \pm \sqrt{\mu^2(1 + \eta_2^2\theta^2) - 1}\right)/(1 - \mu^2)$, and the phase is given by $e^{4i\phi} = (\eta_2\theta - \rho + is)/(\gamma\rho)$, which leads to four values of $\phi$ separated by multiples of $\frac{\pi}{2}$. The existence of these solutions requires $\theta > 0$, and a minimum value of $\mu_0 T$. The "intensity" $\rho$ displays two branches, and it's simple to show via standard linear stability analysis that the lower one is unstable while the upper one is always stable. These solutions never become null, what means that they do not connect with the trivial solution $\psi = 0$ of Eq. (6), which is always stable as in Eq. (3a); in other words, we are in presence of a kind of non-equilibrium first-order transition.

*Experiment.*—In order to give unmistakably evidence of the proposed parametric driving method, next we report our results obtained with a photorefractive oscillator (PRO) in a degenerate FWM configuration. Two counter-propagating beams coming from a single-line laser (Verdi V5, 532 nm) pump a photorefractive strontium-barium-niobate (SBN) crystal contained in a Fabry-Perot cavity. The pump beams form an angle with respect to the cavity axis and thus are not resonated. Above a given threshold (mainly controlled by the crystal orientation with respect to the pump beams) a fraction of the pumps is scattered towards the cavity axis, giving rise to degenerate FWM emission, characterized by the bistability between two phases separated by $\pi$ [47-49,79-81]. A detailed account of the experimental geometry can be found in [79,80]. In order to implement the proposed periodic modulation of the parametric gain, one of the pump beams was reflected by a piezo-mirror before impinging the photorefractive crystal. The piezo-mirror was driven by a function generator and we used a square-wave modulation, finely adjusted in order for the separation between the extreme mirror positions to be exactly half a wavelength, corresponding to a phase jump of $\pi$, i.e. to a pure change of sign of the amplitude of that beam. As a consequence, the FWM parametric gain has a pure sign modulation [57], corresponding to the square-wave case analyzed in the study of Eq. (5). The light leaving the PRO was mixed with a reference beam coming from the same pumping laser in order to perform interferometric measurements leading to the detection of the amplitude and phase of the signal [57,80].

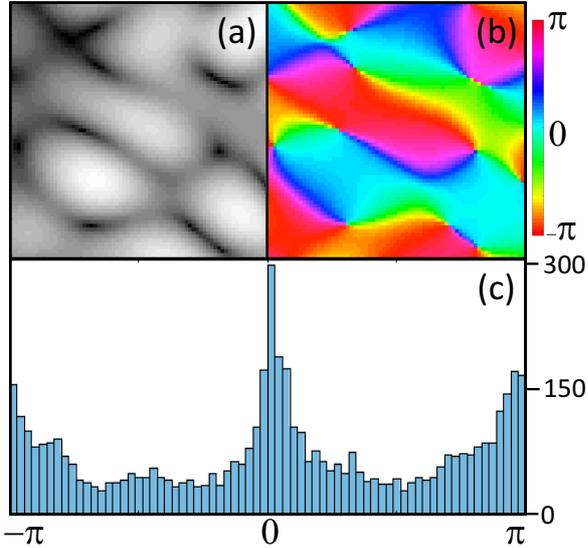

FIG. 2. Phase-bistable behavior in a PRO experiment. (a) is the image of the intensity measured in a 70×70 pixel section beam, showing domains. (b) is its phase profile where is clear that the domains have only two possible phases as we can see in (c), the phase histogram of the above section.

As we are using a spatially extended system (technically, it's a high Fresnel-number optical resonator) different regions can oscillate somehow independently. In particular, in the absence of modulation phase bistability is known to occur, so that different patches of the beam cross section can oscillate in either of the two allowed phases [47-49,80,81]. Figure 2 shows an example of such phase-bistable behavior. When the modulation of the parametric gain is turned on we observe situations like those shown in Fig. 3: four phases emerge in the system, which differ by multiples of π/2.

*Conclusions and outlook.*—We have demonstrated the emergence of phase tetrastability in parametric oscillators, due to the modulation of the driving amplitude. An analogous effect in known to occur in the 4:1 resonance of nonlinear systems close to a Hopf bifurcation [82-88]. In such case the modulation of a parameter at a frequency four times larger than the natural frequency of oscillations is needed. As well as a sufficiently strong nonlinearity is needed for the 4:1 forcing be effective, a case that is not always met; for instance, in the optical domain, such possibility is remote. On the other hand, we are exploiting the usual 2:1 parametric resonance, which is known to be the strongest one, thus favoring the observation of the phase tetrastability. We hope that the phenomena put forward in this Letter can be applied to (quantum) information processing and simulation, metrology and sensing. Work in those directions will be the object of future research.

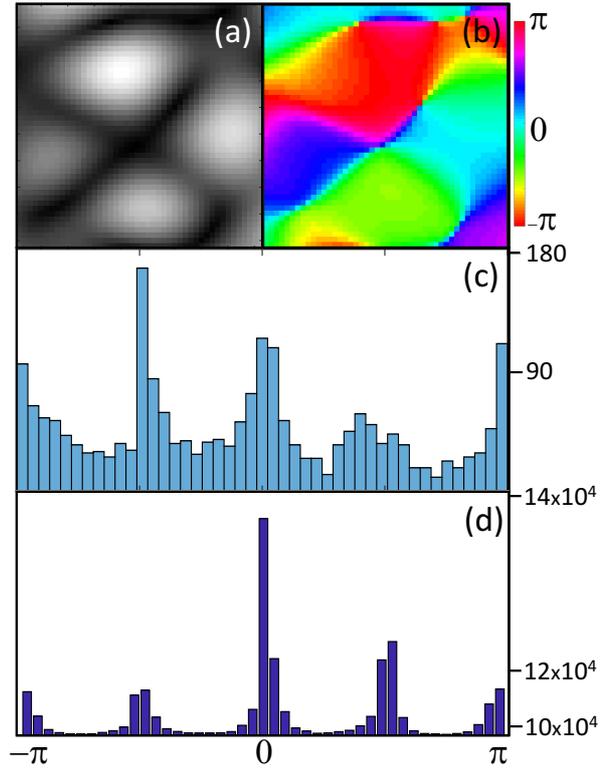

FIG. 3. Phase tetrastability behavior in the Fig. 2 PRO experiment forcing with a 30 Hz rocking in the bidirectional pump. As in Fig 2, (a) is the image of the intensity measured in a 45×45 pixel beam section. (b) is its phase profile where now the domains have four possible phases as we can see in (c), the phase histogram. In (d) we show the phase difference histogram between different pixels in a exponencial scale axis.

We thank Javier García and Martín Sanz for their valuable advice on some techniques used in the experiment. We acknowledge fruitful discussions with Eugenio Roldán and Kestutis Staliunas and also their participation on the early stages of this work. This work was supported by the Spanish Ministerio de Economía, Industria y Competitividad, and FEDER through projects FIS2014-60715-P and FIS2015-65998-C2-1-P.